\documentclass[%
 reprint,
superscriptaddress,
nofootinbib,
 amsmath,amssymb,
 aps,
prd,
]{revtex4-2}

\usepackage{graphicx}
\usepackage{dcolumn}
\usepackage{bm}
\usepackage{hyperref}
\hypersetup{colorlinks,linkcolor=blue,citecolor=green,urlcolor=red}

\usepackage[caption=false]{subfig}

\begin{document}


\title{Analog Hawking radiation emitted by a perfectly reflecting mirror}

\author{Kuan-Nan Lin }
\email{knlinphy@gmail.com}
\affiliation{Asia Pacific Center for Theoretical Physics (APCTP), Pohang, 37673, Korea}
\author{Pisin Chen}
\email{pisinchen@phys.ntu.edu.tw}
\affiliation{Department of Physics and Center for Theoretical Sciences, National Taiwan University, Taipei 10617, Taiwan}
\affiliation{Leung Center for Cosmology and Particle Astrophysics, National Taiwan University, Taipei 10617, Taiwan}

\begin{abstract}
Analog Hawking radiation emitted by a perfectly reflecting mirror in (1+3)-dimensional flat spacetime is investigated. This is accomplished by studying the reflected frequency and momentum based on Einstein's mirror, instead of the canonical way of solving, if possible, wave equations subjected to a dynamical Dirichlet boundary condition. In the case of a finite-size mirror, diffraction pattern appears in the radiation spectrum. Based on the relevant parameters in the proposed Analog Black Hole Evaporation via Lasers experiment, in which the Hawking temperature $T_{H}\simeq 0.03$ eV and the mirror area $\mathcal{A}\simeq (50\;\mu\mathrm{m})^{2}$, the Hawking photon yield is estimated to be $N\simeq 16$/laser shot.
\end{abstract}

\maketitle


\section{\label{Introduction}Introduction}

Since the discovery of Hawking radiation \cite{Hawking1975} in curved spacetime in the 1970s, most of the works of mimicking Hawking radiation via mirror-induced radiation (MIR) in flat spacetime have been devoted to (1+1) dimensions \cite{Moore1970,DeWitt1975,Fulling1976,Davies1977,Carlitz1987,Foo2020,Good2020dS,Good2020ern,Good2020rn,Good2017}. This is because Hawking radiation only propagates radially in the spacetime with a black hole. However, what is experienced in a laboratory is a (1+3)-dimensional flat spacetime, and these additional degrees of freedom complicate the situation of how the vacuum fluctuations interact with a non-point-like mirror. The intent of this paper is to fill in this gap in the literature.

The essence of Hawking radiation in curved spacetime is the gravitational red shift of the wave mode that propagates to the future infinity. In the flying mirror case, the red shift of wave modes is accomplished by the Doppler effect, which is induced upon reflection off the mirror. Thus, it is essential to study how the wave modes are reflected by a flying mirror.


The canonical way of dealing with the flying mirror model is to impose a dynamical Dirichlet boundary condition on a scalar field, whose excitation leads to the mirror-induced radiation. This can be exactly solved in (1+1) dimensions due to the conformally flat property. However, to our awareness, there is no general exact way to solve the problem in spacetime dimension other than (1+1) when the mirror is relativistic. In (1+3)D, however, what have been done include: perfectly reflecting infinite area plane mirror with non-relativistic trajectories \cite{Ford1982,Neto1996,Rego2013}, perfectly reflecting infinite area plane mirror with uniform proper acceleration \cite{candelas1977vacuum}, and semitransparent finite area and thickness mirror with generic relativistic trajectories \cite{Lin2020,Lin2021}.

The necessity of extensions beyond the standard mirror-black hole correspondence is required by laboratory considerations. In particular, the recent Analog Black Hole Evaporation via Lasers (AnaBHEL) Collaboration \cite{AnaBHEL} proposed to generate a flying mirror based on the Chen-Mourou \cite{Chen2017prl,Chen2020} proposal, in which the mirror would have low reflectivity and finite area and thickness, to investigate analog Hawking radiation and the quantum entanglement with its entangled partner particle.

In this paper, instead of working with the conventional approach, we begin by studying the reflected frequency and momentum, which can be derived according to Einstein's special theory of relativity \cite{Einstein1905}, and, from which, we may argue to estimate the lower bound of the number of analog Hawking particles by only considering the normal incident modes, and eventually obtain the quantum radiation spectrum in higher spacetime dimensions by taking the reflected wave mode's Fourier component. In fact, using the Fourier component to study the radiation spectrum of particle creation is quite universal, e.g., in the typical quantum field theory in flat/curved spacetime (including Unruh/Hawking effect), and MIR for perfect/semitransparent mirror, etc.

This paper is organized as follows. In Sec.~\ref{Reflection in (1+1)D}, we begin by reviewing reflections from Einstein's inertial mirror in (1+1)D and, from which, generalization to an accelerated mirror is made. In Sec.~\ref{Reflection in (1+3)D}, we review several crucial aspects, which do not appear in (1+1)D, of reflection by a mirror in (1+3)D. In Sec.~\ref{Quantum particle spectrum}, we first make connection of a general wave mode's Fourier components to the Bogoliubov coefficients in various cases, and later focus on the mirror-induced radiation. Particularly, the radiation spectrum emitted by a (in)finite-size plane mirror, which shows diffraction patterns, in (1+3)D is obtained. In addition, estimation of event yield based on the proposed experiment is made. Conclusion is given in Sec.~\ref{Conclusion}.

\textbf{Notation}: we use the mostly plus metric convention, $G=\hbar=c=k_{B}=1$, and $\mathbb{R}^{1}=(-\infty,+\infty)$.

\section{Reflection in (1+1)D}
\label{Reflection in (1+1)D}

Suppose, according to a static observer in the lab frame $(t,x)$, there is a plane wave $\phi_{inc}=\text{exp}\left(-i\Omega t+iPx\right)$, where $\Omega=|P|>0$ is frequency and $P$ is momentum, that incidents an inertial moving mirror with trajectory: $X(T)=\beta T+x_{0}$, where $\beta=\dot{X}(T)=$ const. is its velocity and $x_{0}=\mathrm{const.}$ According to Einstein's theory of special relativity \cite{Einstein1905}, upon reflection, the incident plane wave will undergo Doppler effect and this reflected wave, as seen by the static observer in the lab frame, is $\phi_{ref}=\text{exp}\left(-i\omega (t-t_{*})+ip(x-x_{*})\right)$, where
\begin{equation}
    \omega=\Omega\left(\frac{1\pm\beta}{1\mp\beta}\right),\quad p=-P\left(\frac{1\pm\beta}{1\mp\beta}\right),
\end{equation}
are, respectively, the Doppler shifted frequency and momentum of the reflected wave (the upper sign is valid for $P<0$ and the lower sign for $P>0$), and $(t_{*},x_{*})$ are some fixed point. For a perfectly reflecting mirror, the wave satisfies time-dependent Dirichlet boundary condition at the mirror: $(t,x)=(T,X(T))$, which leads to
\begin{equation}\label{x*-uniform}
    x_{\mp}^{*}
    =
    t_{*}
    \mp
    x_{*}
    =
    \mp\frac{2}{1\pm \beta}x_{0}.
\end{equation}

A first attempt to generalize the above discussions to an accelerated mirror can be made by giving the above mirror-related constant quantities time dependence. For a right/left-moving reflected wave, $p=\pm\omega$, the reflected waves become
\begin{equation}
    \phi_{ref}(t,x)=\text{exp}\left(-i\int_{x_{\mp}^{*}}^{x_{\mp}(t,x)} d\tilde{x}_{\mp}\;\omega(\tilde{x}_{\mp})\right),
    \label{ref-1}
\end{equation}
where $x_{\mp}(t,x)=t\mp x$ are the light cone coordinates and along a given right/left-moving wave, $x_{\mp}=t\mp x=T\mp X(T)=\mathrm{const.}$, where $(T,X(T))$ is the intersection of the wave and the mirror and $T=T(x_{\mp})$ can be solved as a function of $x_{\mp}$,
\begin{equation}
\begin{aligned}
   \omega(x_{\mp})
   &=
   \Omega
    \left(
    \frac{1\pm\beta(T(x_{\mp}))}{1\mp\beta(T(x_{\mp}))}
    \right),
    \\
    p(x_{\mp})&=-P\left(
    \frac{1\pm\beta(T(x_{\mp}))}{1\mp\beta(T(x_{\mp}))}
    \right),
    \label{ref-wp-1}
\end{aligned}
\end{equation}
are, respectively, the reflected frequency and momentum, which are $T$-dependent (or $(t,x)$-dependent) due to the velocity $\beta(T(x_{\mp}))$. Thus, reflection by the mirror at different time $T$ leads to different Doppler shift.

Since on the worldline of a given reflected wave, $x_{\mp}=T\mp X(T)$, $dx_{\mp}=\left(1\mp\beta(T)\right)dT$ and one obtains
\begin{equation}
    \phi_{ref}(t,x)=\text{exp}\left(-i\Omega\int_{T_{*}}^{T(x_{\mp})} d\tilde{T}\left(1\pm\beta(\tilde{T})\right)\right),
\end{equation}
where $x_{\mp}^{*}=t_{*}\mp x_{*}=T_{*}\mp X(T_{*})$.

In terms of the mirror's proper time $\tau$, which is related to $T$ by the Lorentz gamma $\gamma^{-1}(T)=\sqrt{1-\beta^2(T)}$ by $d\tau = \gamma^{-2}(T)dT$, one obtains
\begin{equation}
    \phi_{ref}(t,x)=\text{exp}\left(-i\Omega\int_{\tau_{*}}^{\tau} d\tilde{\tau}\gamma^{2}\left(1\pm\beta\right)\right).
\end{equation}

Alternatively, one may also begin by solving the (1+1)-dimensional wave equation: $\Box\phi(t,x)=0$, where $\Box=\partial^{\mu}\partial_{\mu}$ is the d'Alembertian operator, with the dynamical Dirichlet boundary condition: $\phi(T,X(T))=0$, where $X(T)$ is the mirror's arbitrary trajectory \cite{Moore1970,DeWitt1975,Fulling1976,Davies1977}. Suppose $\phi=\phi_{inc}-\phi_{ref}$, then for an incident plane wave $\phi_{inc}=\text{exp}\left(-i\Omega t+iPx\right)$, the reflected wave is solved to be $\phi_{ref}=\text{exp}\left(-i\Omega \eta_{\mp}(x_{\mp})\right)$, where $\eta_{\mp}(x_{\mp})=T(x_{\mp})\pm X(T(x_{\mp}))$ is the ray tracing function, for $P<0$ (upper sign) and $P>0$ (lower sign). Noting that $\eta_{\mp}(x_{\mp})=\int_{x_{\mp}^{*}}^{x_{\mp}} d\tilde{x}_{\mp}\partial_{\mp}\eta_{\mp}(\tilde{x}_{\mp})$, where $\eta_{\mp}(x_{\mp}^{*})=0$ determines $x_{\mp}^{*}$ for arbitrary trajectories (\textit{cf.} Eq.~\eqref{x*-uniform}), the reflected wave can also be expressed as
\begin{equation}
    \phi_{ref}(t,x)=\text{exp}\left(-i\Omega\int_{x_{\mp}^{*}}^{x_{\mp}} d\tilde{x}_{\mp} \partial_{\mp}\eta(\tilde{x}_{\mp}) \right),
    \label{ref-exact-1}
\end{equation}
which is identical to the previous generalization, Eq.~\eqref{ref-1}, by identifying the time derivative of the ray tracing function as the reflected frequency, i.e., $\omega(x_{\mp})=\Omega\partial_{\mp}\eta(x_{\mp})$.

\begin{figure}[!htb]
\minipage{0.4\textwidth}
  \includegraphics[width=0.8\linewidth]{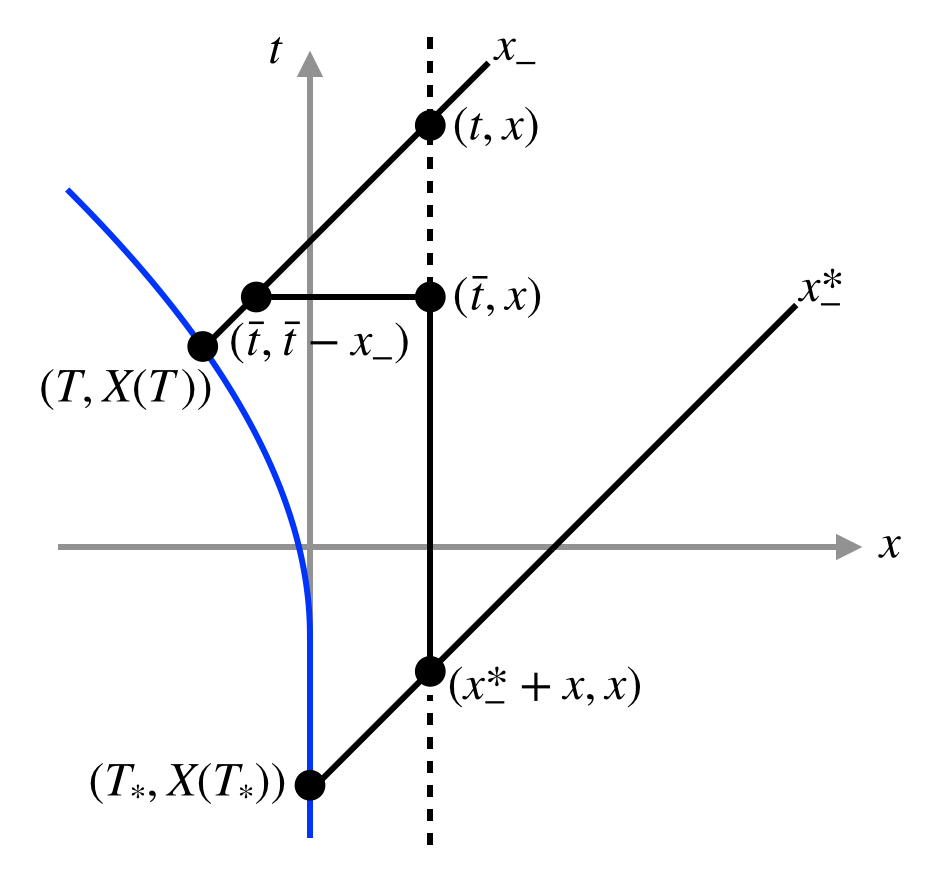}
  \caption{Contours of integration that integrate each wave modes reflected by the mirror starting from $x_{-}^{*}$ to $x_{-}$ once. Blue: mirror's worldline.}
  \label{contour}
\endminipage\hfill
\end{figure}

Although it is possible to express the exponent of the wave reflected by a mirror with generic trajectories in terms of $x_{\mp}$, $T$, or $\tau$ in straightforward manners, caution is required when one wants to express the exponent in terms of $t$ and $x$ since $\beta(T(x_{\mp}))$ depends on $t$ and $x$ simultaneously. For example, if one regards Eq.~\eqref{ref-exact-1} as summing over distinct reflected waves starting from $x_{\mp}^{*}$ to $x_{\mp}$ in the light cone coordinates, and using the fact that the waves after reflection simply propagates to the infinity without any other interactions, then one may write 
\begin{equation}
\begin{aligned}
    \phi_{ref}(t,x)
    &=
    \mathrm{exp}
    \Big(
    -i
    \int_{x_{\mp}^{*}\pm x}^{\bar{t}}d\tilde{t}\;\omega(\tilde{t}\mp x)
    \\
    &\quad
    \pm 
    i
    \int_{x}^{\pm \bar{t}\mp x_{\mp}}d\tilde{x}\;\omega(\bar{t}\mp \tilde{x})
    \Big),
\end{aligned}
\end{equation}
where $\bar{t}$ is an auxiliary time and, e.g., for right-moving reflected waves, the integration contour begins from $(x_{-}^{*}+x,x)$ up to $(\bar{t},x)$ in the vertical direction and then to $(\bar{t},\bar{t}-x_{-})$ in the horizontal direction (see Fig.~\ref{contour}).

These are the generalizations to an accelerated mirror in terms of the observation point $(t,x)$, light cone coordinates $(x_{+},x_{-})$, time $T$, and mirror's proper time $\tau$ based on observations from the results of an inertial mirror.

\section{Reflection in (1+3)D}
\label{Reflection in (1+3)D}

In higher dimensional spacetimes, it is known, if possible, to be difficult to solve the wave equation: $\Box\phi(t,\mathbf{x})=0$ with the boundary condition: $\phi(T,X(T),\mathbf{x}_{\perp})=0$ since conformal symmetry no longer holds. In this section, instead of trying to solve the boundary value problem from first principle, we review the reflection by Einstein's mirror in special relativity, from which we will argue that the consideration of the normal incident modes alone is sufficient to provide a lower bound of the number of analog Hawking radiation emitted by the perfect mirror.

Suppose a plane wave $\phi_{inc}=\text{exp}(-i\Omega t+i\mathbf{P}\cdot\mathbf{x})$ incidents on a perfectly reflecting plane mirror moving at a constant velocity $\beta$, the reflected wave can be derived using Lorentz transformation\footnote{See, e.g., \cite{AA2008} for an alternative derivation based on purely geometric considerations without using Lorentz transformation.} and the result is $ \phi_{ref}=\text{exp}\left(-i\omega t+i\mathbf{p}\cdot\mathbf{x}\right)$, where
\begin{align}\label{reflect3d-w}
    \omega&=\Omega\left(\frac{1\pm 2\beta\cos\theta_{i}+\beta^2}{1-\beta^2}\right),
    \\
    \label{reflect3d-p}
    p_{x}&=-P_{x}\left(\frac{1\pm 2\beta\sec\theta_{i}+\beta^2}{1-\beta^2}\right),\quad \mathbf{p}_{\perp}=\mathbf{P}_{\perp},
\end{align}
where $\theta_{i}$ is the incident angle observed in the lab frame, $P_{x}=\mp \Omega\cos\theta_{i}$, and the reflected angle observed in the lab frame, $\theta_{r}=\cos^{-1}(\pm p_{x}/\omega)$, is given by
\begin{equation}
    \cos\theta_{r}=\frac{(1+\beta^2)\cos\theta_{i}\pm 2\beta}{1\pm 2\beta\cos\theta_{i} +\beta^2}.
    \label{ref-angle}
\end{equation}

The incident frequency $\omega'$ and angle $\theta'$ in the mirror's rest frame are related to the incident frequency $\Omega$ and angle $\theta_{i}$ in the lab frame by
\begin{align}
    \omega'=\Omega\left(\frac{1\pm\beta\cos\theta_{i}}{\sqrt{1-\beta^2}}\right),
    \quad
    \cos\theta'=\frac{\cos\theta_{i}\pm\beta}{1\pm\beta\cos\theta_{i}}.
\end{align}

Since $\cos\theta'$ should be non-negative, it leads to the constraint: $\cos\theta_{i}>|\beta|$, which is also the condition for frequency red shifting, for a receding mirror. Thus, $\theta_{i}$ can be no greater than $\theta_{m}=\cos^{-1}|\beta|$. Physically, this means that waves with incident angles larger than $\theta_{m}$ are not able to catch up the receding mirror. In addition, when $\theta_{i}=\theta_{m}$, $\theta_{r}=-\theta_{m}+\pi$. This shows that the reflected wave can propagate in the same longitudinal direction as the receding mirror but still on the same side as the incident wave in the lab frame. This is known as the \textit{forward reflection}.

In fact, the critical incident angle ($\theta_{i}=\theta_{c}$) beyond which the waves are forwardly reflected can be found by setting $\theta_{r}=\pi/2$ and solve for $\theta_{i}$ in Eq.~\eqref{ref-angle}:
\begin{equation}
    \cos\theta_{i}\left.\right|_{\theta_{i}=\theta_{c}}=\cos\theta_{c}=\mp\left(\frac{2\beta}{1+\beta^2}\right),
\end{equation}
which corresponds to $\theta'=\cos^{-1}|\beta|$ in the rest frame.

In summary, for a receding mirror and in the lab frame, reflected waves are on the same side as the incident waves; however, incident waves with $\theta_{i}\in[0,\theta_{c})$ propagate in the opposite longitudinal direction as the mirror after the reflection; waves with $\theta_{i}\in(\theta_{c},\theta_{m})$ propagate in the same longitudinal direction as the mirror after the reflection, and waves with $\theta_{i}\in(\theta_{m},\pi/2]$ are not able to hit the mirror. On the other hand, in the mirror's rest frame, waves with $\theta_{i}\in[0,\theta_{m})$ propagate in the opposite longitudinal direction as the mirror after the reflection.

For an ultra-relativistically receding mirror, $\beta\simeq \mp 1\pm \delta \beta$, $\delta \beta\ll 1$, and in the small incident angle limit, $\theta_{i}\ll 1$, the leading order reflected frequency and longitudinal momentum are
\begin{align}
    \frac{\omega}{\Omega}&\simeq \frac{\delta \beta}{2}+\frac{\theta_{i}^2}{2\delta\beta},\quad \frac{p_{x}}{\Omega}\simeq \pm\frac{\delta \beta}{2}\mp\frac{\theta_{i}^2}{2\delta\beta}.
\end{align}

If $\theta_{i}\ll \delta\beta$, then the wave mode is Doppler red shifted since $\omega/\Omega\simeq \delta\beta/2\ll 1$; if $1\gg \theta_{i}\gg \delta\beta$, then $\omega/\Omega\simeq \theta_{i}^{2}/2\delta\beta$ and one may naively conclude that the wave mode may get blue shifted if $1\gg \theta_{i}\gg \sqrt{2\delta\beta}$. However, by recalling that $\theta_{c}\simeq \delta\beta\ll 1$ and $\theta_{m}\simeq \sqrt{2\delta\beta}\ll 1$, one notices that the blue shift is in fact excluded. Therefore, for both $\theta_{i}\in [0,\theta_{c}]$ (backward reflection) and $\theta_{i}\in [\theta_{c},\theta_{m}]$ (forward reflection), the wave mode always experiences red shift by an ultra-relativistically receding mirror.

For completeness, when an ultra-relativistically approaching mirror is considered, $\beta\simeq\pm1\mp\delta\beta$, $\delta\beta\ll 1$, the leading order reflected frequency and momentum are
\begin{align}
    \frac{\omega}{\Omega}&\simeq \frac{1+\cos\theta_{i}}{\delta \beta}\gg 1,\quad \frac{p_{x}}{\Omega}\simeq \pm\left(\frac{1+\cos\theta_{i}}{\delta \beta}\right),
\end{align}
where $\theta_{i}\in [0,\pi/2]$. This indicates that the frequency is always Doppler blue shifted, and the reflected wave is highly longitudinal, i.e., $\omega\simeq |p_{x}|\gg |\mathbf{p}_{\perp}|$ for a wide range of incident angle.

Although the reflections in Eqs.~\eqref{reflect3d-w} and \eqref{reflect3d-p} are derived under the assumption of uniform velocity, they in fact also apply for accelerated motion, since the velocity essentially depends on the instantaneous local time $T$, which either corresponds to the retarded time or the advanced time: $t-T=\pm |\mathbf{x}-\mathbf{X}(T)|$, where $(t,\mathbf{x})$ is the observation point and $(T,\mathbf{X}(T))$ is the mirror's trajectory. However, the phase of the reflected wave mode is highly nontrivial since it involves integration over time, which accumulates the effects during acceleration and we don't yet have tools to keep track of the phase's evolution. The strategy in the next section is then to argue that the consideration of the normal incident modes alone (based on Eqs.~\eqref{reflect3d-w} and \eqref{reflect3d-p}) is sufficient to estimate the number of analog Hawking radiation created from vacuum.


\section{Quantum particle spectrum}
\label{Quantum particle spectrum}

In the usual context of quantum field theory in (1+$d$)-dimensional curved spacetime or quantum field theory in (1+$d$)-dimensional flat spacetime in the presence of a suitable external source, the quantum field, say, a massless real scalar field $\phi(t,\mathbf{x})$, can be expanded in terms of the \textit{in} mode $u_{\mathbf{P}}(t,\mathbf{x})$ or the \textit{out} mode $v_{\mathbf{p}}(t,\mathbf{x})$ by
\begin{align}
    \hat{\phi}(t,\mathbf{x})&=\int d^dP\left[\hat{a}_{\mathbf{P}}u_{\mathbf{P}}(t,\mathbf{x})+h.c.\right],
    \\
    &=\int d^dp\left[\hat{b}_{\mathbf{p}}v_{\mathbf{p}}(t,\mathbf{x})+h.c.\right],
\end{align}
where $\Omega=|\mathbf{P}|$, $\omega=|\mathbf{p}|$, $(\hat{a}_{\mathbf{P}},\hat{b}_{\mathbf{p}})$ are annihilation operators, $h.c.$ denotes Hermitian conjugate, and the two mode bases are related by the Bogoliubov transformations
\begin{align}
    v_{\mathbf{p}}(t,\mathbf{x})&=\int d^dP\left[\alpha_{\mathbf{pP}}u_{\mathbf{P}}(t,\mathbf{x})+\beta_{\mathbf{pP}}\bar{u}_{\mathbf{P}}(t,\mathbf{x}) \right],
    \\
     u_{\mathbf{P}}(t,\mathbf{x})&=\int d^dp\left[\bar{\alpha}_{\mathbf{pP}}v_{\mathbf{p}}(t,\mathbf{x})-\beta_{\mathbf{pP}}\bar{v}_{\mathbf{p}}(t,\mathbf{x}) \right],
\end{align}
where $\alpha$ and $\beta$ are the Bogoliubov coefficients and the overbar refers to taking complex conjugation.

In the context of quantum field theory with asymptotic \textit{in} and \textit{out} regions within a given coordinate system, whether in flat spacetime, e.g., particle creation by a moving mirror, or in curved spacetime, e.g., cosmological particle production, one often encounters
\begin{align}
   \lim_{t\rightarrow-\infty} u_{\mathbf{P}}(t,\mathbf{x})
   &=
   \frac{1}{(2\pi)^{3/2}\sqrt{2\Omega}}\text{exp}\left(-i\Omega t+i\mathbf{P}\cdot\mathbf{x}\right),
   \\
    \lim_{t\rightarrow+\infty} v_{\mathbf{p}}(t,\mathbf{x})
    &=
    \frac{1}{(2\pi)^{3/2}\sqrt{2\omega}}\text{exp}\left(-i\omega t+i\mathbf{p}\cdot\mathbf{x}\right),
\end{align}
so by taking the Fourier transformation of the \textit{out} mode at the infinite past, i.e.,
\begin{align}
    \lim_{t\rightarrow-\infty}v_{\mathbf{p}}(t,\mathbf{x})=\int &\frac{d^3P}{(2\pi)^{3/2}\sqrt{2\Omega}}\biggl[e^{-i\Omega t+i\mathbf{P}\cdot\mathbf{x}}\frac{\tilde{v}_{\mathbf{p}}(\Omega,\mathbf{P})}{\sqrt{2\Omega}}
    \nonumber
    \\
    &+e^{i\Omega t-i\mathbf{P}\cdot\mathbf{x}}\frac{\tilde{v}_{\mathbf{p}}(-\Omega,-\mathbf{P})}{\sqrt{2\Omega}}\biggr],
\end{align}
allows us to identify the Bogoliubov coefficients as
\begin{equation}
    \alpha_{\mathbf{pP}}=\frac{\tilde{v}_{\mathbf{p}}(\Omega,\mathbf{P})}{\sqrt{2\Omega}},\quad 
    \beta_{\mathbf{pP}}=\frac{\tilde{v}_{\mathbf{p}}(-\Omega,-\mathbf{P})}{\sqrt{2\Omega}}.
    \label{Bogoliubov-FT-1}
\end{equation}

On the other hand, in non-dynamical situations, e.g., Unruh effect in (1+1) dimensions, which involves two different coordinate systems, one encounters
\begin{align}
    v_{\omega}(t,x)&=\frac{1}{(2\pi)^{1/2}\sqrt{2\omega}}\text{exp}\left(-i\omega t+i\omega x\right),
    \\
    u_{\Omega}(\tau,\xi)&=\frac{1}{(2\pi)^{1/2}\sqrt{2\Omega}}\text{exp}\left(-i\Omega \tau+i\Omega \xi\right),
\end{align}
where $(t,x)$ are Minkowski coordinates and $(\tau,\xi)$ are Rindler coordinates, which are related by the transformation: $t=a^{-1}e^{a\xi}\sinh(a\tau)$, $x=a^{-1}e^{a\xi}\cosh(a\tau)$, $a>0$. On the worldline of a Rindler observer at $\xi=0$, the Minkowski mode can be expanded in terms of the Rindler mode by
\begin{align}
    v_{\omega}(\tau)&=\frac{1}{\sqrt{4\pi\omega}}\cdot\text{exp}\left(\frac{i\omega}{a}e^{-a\tau}\right)
    \nonumber
    \\
    &=\int_{0}^{\infty}\frac{d\Omega}{\sqrt{2\pi}\sqrt{2\Omega}}\left[e^{-i\Omega\tau}\frac{\tilde{v}_{\omega}(\Omega)}{\sqrt{2\Omega}}+e^{i\Omega\tau}\frac{\tilde{v}_{\omega}(-\Omega)}{\sqrt{2\Omega}}\right],
\end{align}
which allows the identifications:
\begin{equation}
    \alpha_{\omega\Omega}=\frac{\tilde{v}_{\omega}(\Omega)}{\sqrt{2\Omega}},\quad 
    \beta_{\omega\Omega}=\frac{\tilde{v}_{\omega}(-\Omega)}{\sqrt{2\Omega}},
\end{equation}
where $\tilde{v}_{\omega}(\Omega)=\sqrt{2\Omega^{2}/\omega}(-i\omega/a)^{i\Omega/a}\Gamma(-i\Omega/a)/(2\pi a)$.

In this paper, we are interested in particle creation by a relativistic perfectly reflecting mirror, thus only the former dynamical situation will be relevant to our following discussions.

\subsection{Flying mirror in (1+1)D}

\begin{figure}[!htb]
\minipage{0.4\textwidth}
  \includegraphics[width=0.8\linewidth]{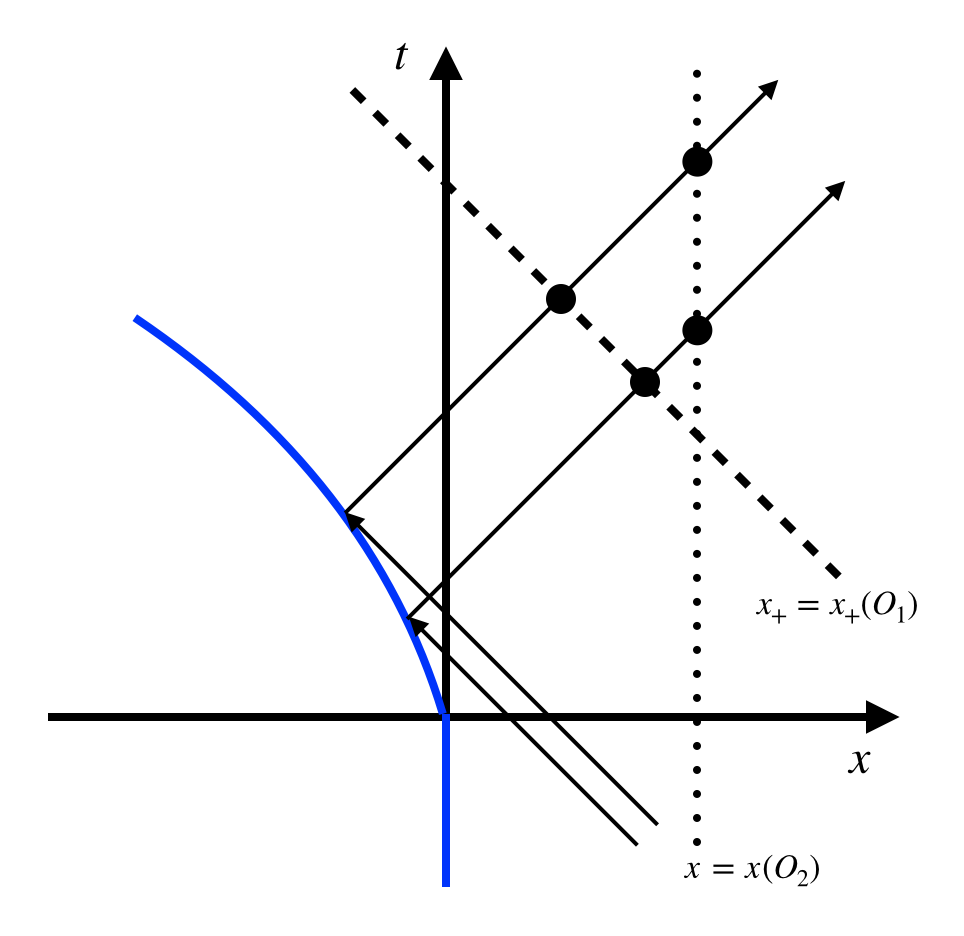}
  \caption{Reflection of plane waves by a relativistically receding perfect mirror in (1+1)-dimensions. Blue: mirror's trajectory. Dashed line: observer $\mathcal{O}_{1}$'s worldline $x_{+}=x_{+}(\mathcal{O}_{1})$. Dotted line: observer $\mathcal{O}_{2}$'s worldline $x=x(\mathcal{O}_{2})$.}
  \label{trajectory-DF-1}
\endminipage\hfill
\end{figure}

To mimic Hawking radiation emitted by a black hole formed from gravitational collapse, we shall consider a perfect mirror whose trajectory asymptotes the Davies-Fulling one: $X(T)\sim-T-A\text{exp}(-2\kappa T)-B,\left\{A,B,\kappa\right\}>0$ at late times \cite{Davies1977} (see Fig.~\ref{trajectory-DF-1}).

Since the \textit{out} mode $v_{\omega}(x_{+}(\mathcal{O}_{1}))\sim\text{exp}(-i\omega x_{-})$ is a plane wave basis (up to a normalization factor) for an \textit{out} observer $\mathcal{O}_{1}$ at $x_{+}=x_{+}(\mathcal{O}_{1})$, the reflected waves, which have experienced Doppler red shift, that reach $x_{+}=x_{+}(\mathcal{O}_{1})$ can be expanded in terms of $v_{\omega}(x_{+}(\mathcal{O}_{1}))$, which further leads to the interpretation of Fourier modes as Bogoliubov coefficients. In this case, we have $\omega(x_{-})/\Omega\sim  A\kappa\cdot \text{exp}(-2\kappa T(x_{-}))\sim  A\kappa\cdot \text{exp}(-\kappa x_{-})$, and the Fourier transformation of the \textit{in} mode at $x_{+}(\mathcal{O}_{1})$:
\begin{align}
    &\tilde{u}_{\Omega}(\omega)=\frac{\sqrt{2\omega^2}}{\sqrt{\pi}}\int_{\mathbb{R}^{1}} dx_{-}e^{i\omega x_{-}}u_{\Omega}(x_{+}(\mathcal{O}_{1}))
    \nonumber
    \\
    &=
    -\frac{\sqrt{\omega^2}}{\sqrt{2\pi^{2}\Omega}}\int_{\mathbb{R}^{1}} dx_{-}e^{i\omega x_{-}}\text{exp}\left(-i\int_{x_{-}^{*}}^{x_{-}} d\tilde{x}_{-}\omega(\tilde{x}_{-})\right)
    \nonumber
    \\
    &\simeq -\frac{\sqrt{\omega^2}}{\sqrt{2\pi^{2}\Omega}}\int_{\mathbb{R}^{1}} dx_{-}e^{i\omega x_{-}}\text{exp}\left(i\Omega A e^{-\kappa x_{-}}\right)
    \nonumber
    \\
    &=-\frac{\sqrt{\omega^2}}{\sqrt{2\pi^{2}\Omega}}\frac{(-i\Omega A)^{i\omega/\kappa}}{\kappa}\Gamma\left(\frac{-i\omega}{\kappa}\right),
\end{align}
where the third line extrapolates the asymptotic trajectory to the infinite past to obtain an approximate thermal spectrum \cite{Davies1977}, and the Bogoliubov coefficients are
\begin{equation}
    \bar{\alpha}_{\omega\Omega}=\frac{\tilde{u}_{\Omega}(\omega)}{\sqrt{2\omega}},\quad \beta_{\omega\Omega}=-\frac{\tilde{u}_{\Omega}(-\omega)}{\sqrt{2\omega}}.
\end{equation}

On the other hand, one intuitively expects an observer $\mathcal{O}_{2}$ at $x=x(\mathcal{O}_{2})$ should obtain the same spectrum as that obtained by $\mathcal{O}_{1}$ at $x_{+}=x_{+}(\mathcal{O}_{1})$. Indeed, since $x_{-}=t-x$, the integration over $x_{-}$ in the Fourier transform of $u_{\Omega}(t,x)$ can also be regarded as an integration over the observation time $t$ at a fixed spatial position $x=x(\mathcal{O}_{2})$. In this perspective, we have
\begin{align}
    &\tilde{u}_{\Omega}(\omega)=\frac{\sqrt{2\omega^2}}{\sqrt{\pi}}\int_{\mathbb{R}^{1}} dt \;e^{i\omega x_{-}}u_{\Omega}(x(\mathcal{O}_{2}))
    \nonumber
    \\
    &=-\frac{\sqrt{\omega^2}}{\sqrt{2\pi^{2}\Omega}}\int_{\mathbb{R}^{1}} dt\;e^{i\omega x_{-}}\text{exp}\left(-i\int_{t_{*}}^{t} d\tilde{t}\;\omega(\tilde{t}-x)\right)
    \nonumber
    \\
    &\simeq -\frac{\sqrt{\omega^2}}{\sqrt{2\pi^{2}\Omega}}e^{-i\omega x}\int_{\mathbb{R}^{1}} dt\;e^{i\omega t}\text{exp}\left(i\Omega Ae^{\kappa x} e^{-\kappa t}\right)
    \nonumber
    \\
    &=-\frac{\sqrt{\omega^2}}{\sqrt{2\pi^{2}\Omega}}\frac{(-i\Omega A)^{i\omega/\kappa}}{\kappa}\Gamma\left(\frac{-i\omega}{\kappa}\right),
\end{align}
which is identical to the result of $\mathcal{O}_{1}$. Finally, let us comment that identical result can also be obtained if one integrates along time $T$.

\subsection{Flying mirror in (1+3)D}

In higher spacetime dimensions, the wave mode's non-vanishing transverse momentum significantly increases technical difficulties. As mentioned in the previous section, for a wave mode that passes through the spacetime point $(t,\mathbf{x})$, the intersection of its worldline and the mirror's trajectory $(T,\mathbf{X}(T))$ also depends on the transverse coordinates by $t-T=\pm |\mathbf{x}-\mathbf{X}(T)|$. Therefore, the (1+1)-dimensional light cone coordinates $x_{\mp}=t\mp x=T\mp X(T)$ only apply for wave modes that propagate in the direction normal to the mirror's transverse area.

Although the complete treatment of wave modes with non-vanishing transverse momentum not yet exists, it turns out that we can still estimate the lower bound of the number of particles that are relevant for analog Hawking radiation based on the discussion of reflection in Sec.~\ref{Reflection in (1+3)D}. 

Let us again consider a perfectly reflecting mirror with the asymptotic trajectory: $X(T)\sim-T-A\text{exp}(-2\kappa T)-B,\left\{A,B,\kappa\right\}>0$, and we assume that the mirror has an area transverse to its prescribed motion. According to Sec.~\ref{Reflection in (1+3)D}, the critical angle $\theta_{c}$ and the maximum angle $\theta_{m}$ for the incident \textit{in} modes that are able to catch up and get reflected by the receding mirror are 
\begin{align}\label{angle-hawking-1}
    \theta_{c}(T)&\sim 2A\kappa \;e^{-2\kappa T}+\mathcal{O}(e^{-4\kappa T})\ll 1,
    \\
    \label{angle-hawking-2}
    \theta_{m}(T)&\sim \sqrt{4A\kappa}\;e^{-\kappa T}+\mathcal{O}(e^{-3\kappa T})\ll 1.
\end{align}

As the mirror continues to accelerate, these angles will only get narrower. Since the analog Hawking radiation is only relevant to the extreme redshift due to the asymptotic behavior, Eqs.~\eqref{angle-hawking-1} and \eqref{angle-hawking-2} imply that only wave modes with sufficiently small incident angles contribute. For these \textit{in} modes, the leading order reflected frequency and longitudinal momentum are
\begin{align}
    \frac{\omega}{\Omega}&\sim  A\kappa e^{-2\kappa T}+\frac{\theta_{i}^{2}}{4A\kappa e^{-2\kappa T}},
    \label{ref-freq-3D}
    \\
    \frac{p_{x}}{\Omega}&\sim  A\kappa e^{-2\kappa T}-\frac{\theta_{i}^{2}}{4A\kappa e^{-2\kappa T}}.
\end{align}

According to the discussion in Sec.~\ref{Reflection in (1+3)D}, for \textit{in} modes with $\theta_{i}\in[0,\theta_{c}(T)]$, the reflected wave modes essentially have vanishing transverse momentum to the leading order since $\omega/\Omega=p_{x}/\Omega\sim  A\kappa e^{-2\kappa T}\ll 1$, and the situation is effectively (1+1)-dimensional and the standard Doppler red shift for Hawking radiation is recovered; for $\theta_{i}\in [\theta_{c},\theta_{m}]$, the \textit{in} modes are also red shifted; however, in a different manner by $\omega/\Omega=-p_{x}/\Omega\sim \theta_{i}^{2}/(4A\kappa e^{-2\kappa T})=(\theta_{i}/\theta_{m}(T))^{2}\ll 1$. Since the critical and the maximal angle decrease exponentially in time, as the mirror continues to accelerate, it quickly turns out that only \textit{in} modes with vanishing transverse momenta (normal incident modes) could get reflected by the mirror. Thus, we will focus on the normal incident modes to estimate the lower bound of the number of analog Hawking particles for future experimental purpose.

\begin{figure}[!htb]
\minipage{0.4\textwidth}
  \includegraphics[width=0.8\linewidth]{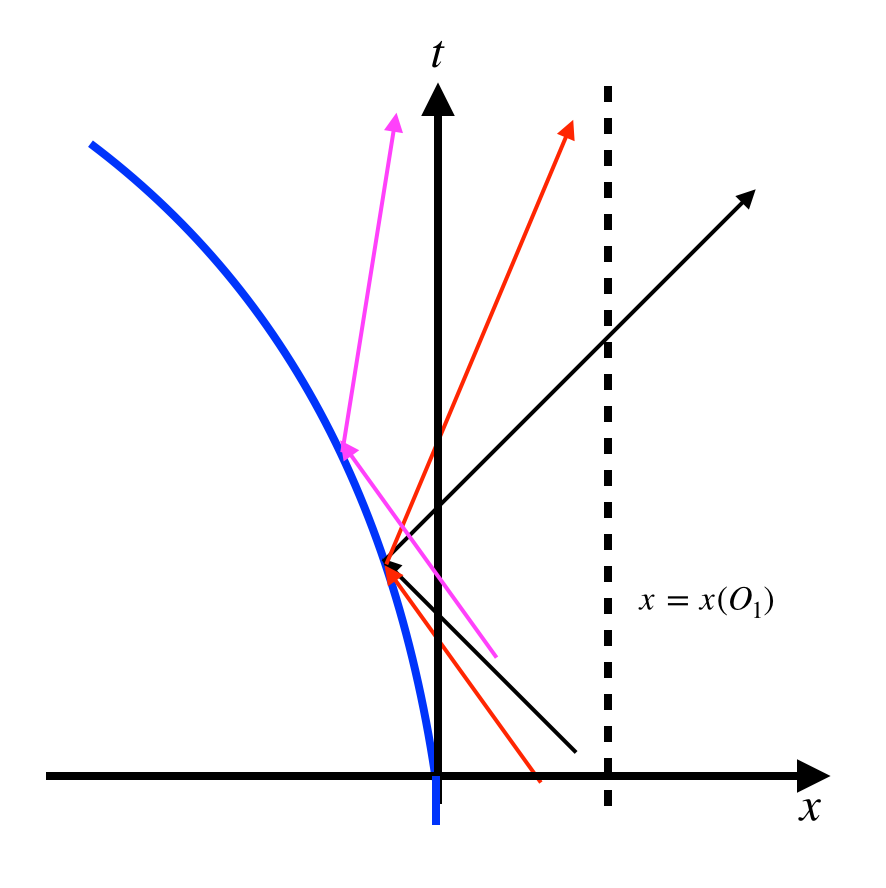}
  \caption{Reflection of plane waves by a relativistically receding perfect mirror in higher dimensions. Blue: mirror's trajectory. Dashed line: observer $\mathcal{O}_{1}$'s worldline $x=x(\mathcal{O}_{1})$.}
  \label{trajectory-DF-2}
\endminipage\hfill
\end{figure}

In this section, since we consider the presence of a mirror in (1+3)-dimensional flat spacetime, one can search for the quantum radiation emitted to either sides of the mirror. In addition, since the analog Hawking radiation belong to the incident wave modes that experience extreme red shift, we will focus on an \textit{out} observer $\mathcal{O}_{1}$ located at $\mathbf{x}=\mathbf{x}(\mathcal{O}_{1})$ (see Fig.~\ref{trajectory-DF-2}), then the Fourier transform of the reflected wave modes by the mirror can be interpreted as the spectrum of radiation observed by $\mathcal{O}_{1}$.

In (1+3)-dimensional flat spacetime, a scalar field operator to an infinite transverse area mirror's right can be generally expanded by
\begin{equation}
    \begin{aligned}
    \hat{\phi}(t,\mathbf{x})&=\int_{\mathbb{R}^{3}}
    d^{3}P\left[\hat{a}_{\mathbf{P}}u_{\mathbf{P}}(t,\mathbf{x})+h.c.\right],
    \end{aligned}
\end{equation}
where $t\in \mathbb{R}^{1},\;\mathbf{x}_{\perp}\in\mathbb{R}^{2}$, $x\geq X$, and 
\begin{equation}
    \lim_{t\rightarrow-\infty}
    u_{\mathbf{P}}(t,\mathbf{x})
    =
    \frac{1}{(2\pi)^{3/2}\sqrt{2\Omega}}\text{exp}\left(-i\Omega t+i\mathbf{P}\cdot\mathbf{x}\right)
\end{equation}
is the \textit{in} mode and $\Omega=|\mathbf{P}|$. However, according to an \textit{out} observer $\mathcal{O}_{1}$ on the mirror's right-hand side, only \textit{in} modes with $P_{x}<0$ are relevant for particle creation since they propagate to the left and can catch up the left-moving mirror and get reflected back (\textit{in} modes with $P_{x}>0$ do not interact with the mirror and simply remain as vacuum fluctuations). In addition, since we are interested in the \textit{in} modes with negligible transverse momenta to estimate the lower bound of particle creation, we can write down $\hat{a}_{\mathbf{P}}=2\pi\delta^{(2)}(\mathbf{P}_{\perp})\Theta(-P_{x})\hat{c}_{\Omega}$, where the commutation relation in (1+3)-dimensions: $[\hat{a}_{\mathbf{P}},\hat{a}_{\mathbf{P}'}^{\dagger}]=\delta^{(3)}(\mathbf{P}-\mathbf{P}')=\delta(\Omega-\Omega')\delta^{(2)}(\mathbf{P}_{\perp})$ implies 
\begin{equation}
    \begin{aligned}
        \left[\hat{c}_{\Omega},\hat{c}_{\Omega'}^{\dagger}\right]=\frac{\delta(\Omega-\Omega')}{\mathcal{A}_{\infty}},
    \end{aligned}
\end{equation}
where the mirror's infinite area $\mathcal{A}_{\infty}$ appears due to
\begin{equation}
    \delta^{(2)}(\mathbf{P}_{\perp}=\mathbf{0})
    =
    \int_{\mathbb{R}^{2}}\frac{d^{2}x_{\perp}}{(2\pi)^{2}}
    =
    \frac{\mathcal{A}_{\infty}}{(2\pi)^{2}}
    \rightarrow
    \infty.
\end{equation}

Thus, the field operator on the infinite area mirror's right-hand side becomes effectively (1+1)-dimensional:
\begin{equation}
    \begin{aligned}
    \hat{\phi}(t,\mathbf{x})&=\int_{\mathbb{R}^{1}}
    d\Omega\left[2\pi\hat{c}_{\Omega}\left.u_{\mathbf{P}}(t,\mathbf{x})\right|_{\mathbf{P}_{\perp}=\mathbf{0}}+h.c.\right],
    \end{aligned}
\end{equation}
where $\Omega=-P_{x}$ and
\begin{align}
    \left.u_{\mathbf{P}}(t,\mathbf{x})\right|_{\mathbf{P}_{\perp}=0}=\frac{e^{-i\Omega x_{+}}-e^{-i\int_{x_{-}^{*}}^{x_{-}} d\tilde{x}_{-}\omega(\tilde{x}_{-})}}{(2\pi)^{3/2}\sqrt{2\Omega}},
\end{align}
where the first term is the left-moving incident wave mode, and the second term is the reflected wave mode with the time-dependent frequency given by Eq.~\eqref{ref-freq-3D} in the asymptotic late time in order to mimic Hawking radiation.

On the other hand, since what an \textit{out} observer receives are right-moving modes $(p_{x}>0)$, which interact with the mirror when tracing back in time, the field operator can also be expanded by
\begin{equation}
    \begin{aligned}
    \hat{\phi}(t,\mathbf{x})&=
    \int_{p_{x}>0} d^{3}p\left[\hat{b}_{\mathbf{p}}v_{\mathbf{p}}(t,\mathbf{x})+h.c.\right],
    \end{aligned}
\end{equation}
where
\begin{equation}
    \lim_{t\rightarrow+\infty} v_{\mathbf{p}}(t,\mathbf{x})
    =
    \frac{1}{(2\pi)^{3/2}\sqrt{2\omega}}\text{exp}\left(-i\omega t+i\mathbf{p}\cdot\mathbf{x}\right).
\end{equation}

According to the discussion in Sec.~\ref{Quantum particle spectrum}, the reflected \textit{in} mode can be decomposed into Fourier components by
\begin{align}
    &u_{\mathbf{P}}(\mathbf{x}(\mathcal{O}_{1}))=-\frac{e^{-i\int_{x_{-}^{*}}^{x_{-}} d\tilde{x}_{-}\omega(\tilde{x}_{-})}}{(2\pi)^{3/2}\sqrt{2\Omega}}
    \label{ref-in-mode-3D}
    \\
    &=\int_{p_{x}>0}\frac{d^{3}p}{(2\pi)^{3/2}\sqrt{2\omega}}\left[e^{ip\cdot x}\frac{\tilde{u}_{\Omega}(\omega,\mathbf{p})}{\sqrt{2\omega}}+e^{-ip\cdot x}\frac{\tilde{u}_{\Omega}(-\omega,-\mathbf{p})}{\sqrt{2\omega}}\right],
    \nonumber
\end{align}
where $\omega=p^{t}=-p_{t}=|\mathbf{p}|$, $p\cdot x=-\omega t+\mathbf{p}\cdot\mathbf{x}$, and the Bogoliubov coefficients are given by
\begin{equation}
\begin{aligned}
    \bar{\alpha}_{\mathbf{pP}}
    &=
    \frac{\tilde{u}_{\Omega}(\omega,\mathbf{p})}{\sqrt{2\omega}}
    =
    \frac{\delta^{(2)}(\mathbf{p}_{\perp})\underline{u}_{\Omega}(\omega,p_{x})}{\sqrt{2\omega}},
    \\
    \beta_{\mathbf{pP}}
    &=
    -\frac{\tilde{u}_{\Omega}(-\omega,-\mathbf{p})}{\sqrt{2\omega}}
    =
    -\frac{\delta^{(2)}(\mathbf{p}_{\perp})\underline{u}_{\Omega}(-\omega,-p_{x})}{\sqrt{2\omega}},
\end{aligned}
\end{equation}
where
\begin{align}
    \underline{u}_{\Omega}(\omega,p_{x})
    &\simeq-\frac{\sqrt{\omega^2}}{\sqrt{2\pi^{2}\Omega}}e^{-i\omega x}\int_{\mathbb{R}^{1}} dt\;e^{i\omega t}\text{exp}\left(i\Omega Ae^{\kappa x} e^{-\kappa t}\right)
    \nonumber
    \\
    &=-\frac{\sqrt{\omega^2}}{\sqrt{2\pi^{2}\Omega}}\frac{(-i\Omega A)^{i\omega/\kappa}}{\kappa}\Gamma\left(\frac{-i\omega}{\kappa}\right),
\end{align}
where we have used Eq.~\eqref{ref-freq-3D} and focused on $\theta_{i}= 0$. Thus, the Bogoliubov coefficients of interest are
\begin{align}
    \bar{\alpha}_{\mathbf{pP}}&\simeq-\frac{\delta^{(2)}(\mathbf{p}_{\perp})}{2\pi\kappa}\frac{\sqrt{\omega}}{\sqrt{\Omega}}(-i\Omega A)^{i\omega/\kappa}\Gamma\left(-\frac{i\omega}{\kappa}\right),
    \\
    \beta_{\mathbf{pP}}&\simeq\frac{\delta^{(2)}(\mathbf{p}_{\perp})}{2\pi\kappa}\frac{\sqrt{\omega}}{\sqrt{\Omega}}(-i\Omega A)^{-i\omega/\kappa}\Gamma\left(\frac{i\omega}{\kappa}\right).
\end{align}

The mean occupation number of analog Hawking particles is thus
\begin{align}
    \frac{d^3N}{d^{3}p}&=\left<0;\text{in}\right|\hat{b}_{\mathbf{p}}^{\dagger}\hat{b}_{\mathbf{p}}\left|0;\text{in}\right>
    \nonumber
    \\
    &=\lim_{\mathbf{p}'\rightarrow\mathbf{p}}\frac{(2\pi)^2}{\mathcal{A}_{\infty}}\int_{0}^{\infty}d\Omega\beta_{\mathbf{pP}}\bar{\beta}_{\mathbf{p'P}}
    \nonumber
    \\
    &\simeq\frac{(2\pi)^2\delta^{(2)}(\mathbf{p}_{\perp})}{\mathcal{A}_{\infty}}\frac{\delta^{(2)}(\mathbf{p}_{\perp})\Delta t}{2\pi}\left(\frac{1}{e^{\omega/T_{H}}-1}\right),
    \label{analog Hawking spectrum}
\end{align}
where the divergence $\Delta t=2\pi\delta(\omega-\omega'=0)\rightarrow\infty$ is due to the extrapolation of the asymptotic behavior to the infinite past $t\rightarrow-\infty$ \cite{Davies1977}, the first fraction in the last line equals to unity for $\mathbf{p}_{\perp}=\mathbf{0}$, and $T_{H}=\kappa/2\pi$ is the analog Hawking temperature. By integrating over the transverse momentum, one obtains
\begin{equation}\label{analog Hawking spectrum-2}
    \frac{1}{\Delta t}\frac{dN}{dp_{x}}
    =
    \frac{1}{\Delta t}\frac{dN}{d\omega}
    \simeq
    \frac{1}{2\pi}\left(\frac{1}{e^{\omega/T_{H}}-1}\right).
\end{equation}

Interestingly, while the distribution \eqref{analog Hawking spectrum} does depend on the mirror's area, the area dependence disappears upon the integration over $\mathbf{p}_{\perp}$. This is because the situation reduces to (1+1)-dimensional when $\mathbf{p}_{\perp}=\mathbf{0}$ is considered, and there is no notion of area since the space is one-dimensional. However, when the transverse momentum is not negligible, then both the radiation spectrum (distribution) and the total number of particles may depend on the mirror's area \cite{Lin2021}.

When a finite transverse area mirror and normal incident \textit{in} modes are considered, the transverse coordinates $\mathbf{x}_{\perp}$ is restricted to the space covered by the mirror's finite transverse area instead of $\mathbf{x}_{\perp}\in\mathbb{R}^{2}$, and Eqs.~\eqref{analog Hawking spectrum} and \eqref{analog Hawking spectrum-2} should be the first order approximation in the large area limit. In the case of a semitransparent mirror, the finite area form factor has been derived by considering detailed local interaction between the scalar field and the mirror \cite{Lin2021}, and it turns out that, for a square mirror with area $\mathcal{A}=L\times L$, one can simply replace the Dirac delta functions in Eq.~\eqref{analog Hawking spectrum} by sinc functions using the identity:
\begin{align}\label{sinc}
    \lim_{L\rightarrow\infty}\text{sinc}(pL/2)=2\pi\delta(p)/L.
\end{align}

Since the form factor is purely geometrical, i.e., it does not depend on the mirror's reflectivity or its motion, we obtain the radiation spectrum by a perfectly reflecting, finite area, square mirror from Eqs.~\eqref{analog Hawking spectrum} and \eqref{sinc} as
\begin{align}
    \frac{d^3N}{d^{3}p}\simeq\frac{\mathcal{A}\Delta t}{(2\pi)^3}\;\text{sinc}^2\left(\frac{p_{y}L}{2}\right)\text{sinc}^2\left(\frac{p_{z}L}{2}\right)\left(\frac{1}{e^{\omega/T_{\mathrm{eff}}}-1}\right).
    \nonumber
\end{align}

Note that the modification in the dynamical part, i.e., $T_{H}\rightarrow T_{\mathrm{eff}}=\kappa/[(1+\cos\theta)\pi]$, is not derived from first principle in the case of a perfect mirror in this paper. However, the experiences in \cite{Lin2020,Lin2021} for a semitransparent mirror indicate that the exponential $e^{\omega/T_{\mathrm{eff}}}$ has a clear physical origin: its appearance is related to the Doppler phase shift in the direction parallel to the mirror's motion. In addition, since the phase shift of the reflected wave mode is only related to the mirror's motion, while the reflectivity only affects the amplitude of the wave mode, we expect the appearance of the effective temperature to be valid.

On the other hand, the appearance of the sinc functions is a manifestation of diffraction due to the mirror's finite transverse geometry. As in standard optics, in the far field, using $p_{y}\sim\omega y/R$ and $p_{z}\sim\omega z/R$, where $R\gg|\mathbf{x}_{\perp}|$ is the longitudinal distance between the observer $\mathcal{O}_{1}$ and the mirror at the instant of emission, the arguments of the sinc functions become $\omega L y/(2R)\sim \omega L\theta\cos\phi/2$ and $\omega L z/(2R)\sim \omega L\theta\sin\phi/2$, which are also justified by $p_{y}=\omega\sin\theta\cos\phi$ and $p_{z}=\omega\sin\theta\sin\phi$ when $\theta\ll 1$, where $\theta$ and $\phi$ are the emission angles.

Similarly, for a circular mirror of diameter $D$, the sinc function is usually replaced by the jinc function, and the radiation spectrum becomes
\begin{align}
    \frac{d^3N}{d^{3}p}\simeq\frac{\mathcal{A}\Delta t}{(2\pi)^3}\left[2\text{jinc}\left(\frac{|\mathbf{p}_{\perp}| D}{2}\right)\right]^2\left(\frac{1}{e^{\omega/T_{\mathrm{eff}}}-1}\right),
    \nonumber
\end{align}
where $\mathcal{A}=D^2\pi/4$, $\text{jinc}(x)=J_{1}(x)/x$, $J_{1}(x)$ is the Bessel function of the first kind of order 1. This leads to the standard Airy pattern in optics for circular mirrors, and the first minimum of the jinc function occurs at $\theta=\theta_{1}=7.66/(\omega D)=1.22\lambda/D$, where $\lambda=2\pi/\omega$ is the wavelength of the emitted analog Hawking particle.

The total occupation number can be obtained by integrating over the momentum $\mathbf{p}$. For $\mathcal{A}\rightarrow\infty$, there is no diffraction, and the above three types of mirrors all lead to the same yield:
\begin{align}
    N&\simeq\frac{\Delta t}{2\pi}\int_{\omega_{1}}^{\omega_{2}} \frac{d\omega}{e^{2\pi\omega/\kappa}-1}
    \nonumber
    \\
    &=\frac{T_{H}\Delta t}{2\pi}\log\left(\frac{1-e^{-\omega_{2}/T_{H}}}{1-e^{-\omega_{1}/T_{H}}}\right),
    \label{total-number}
\end{align}
where $(\omega_{1},\omega_{2})$ is the frequency range of interest.

For a perfect square mirror with finite area $\mathcal{A}$, diffraction emerges. However, analytic expressions of spectra can only be obtained under certain approximations. For simplicity, in the case of a square mirror, the frequency spectrum in the low frequency regime: $\omega\ll L^{-1}$ is
\begin{align}\label{omega-small}
    \frac{dN}{d\omega}\simeq\frac{\mathcal{A}\Delta t}{(2\pi)^2}\frac{\kappa\omega}{\pi}\log\big(1+e^{-\pi\omega/\kappa} \big),
\end{align}
while the angular spectrum with $\theta\ll 1$ is
\begin{align}
    \frac{d^{2}N}{d\Omega}\simeq\frac{\mathcal{A}\Delta t}{(2\pi)^3}\frac{\kappa^3\zeta(3)}{4\pi^3}\Big[1-\frac{\zeta(5)(\kappa L)^2-3\pi^2\zeta(3)}{4\pi^2\zeta(3)}\theta^2 \Big],
    \label{angular-small}
\end{align}
where $d\Omega=\sin\theta d\theta d\phi$ is the differential solid angle (not to be confused with the frequency $\Omega$ that appears in the previous discussions), $\zeta$'s are the Riemann zeta functions. These spectra show the area dependence for both the distribution and the total number of particles $N$. After all, particles created with non-vanishing transverse momenta extend the situation beyond (1+1)-dimensions, so the notion of area should naturally appear.

\begin{figure}[!htb]
\minipage{0.45\textwidth}
  \includegraphics[width=0.8\linewidth]{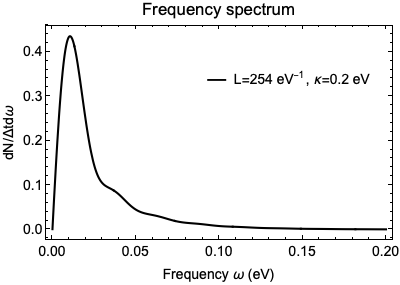}
  \caption{Frequency spectrum per unit time of analog Hawking radiation. The mild bump around $0.04$ eV is not a numerical artifact but a real diffraction effect due to the finite size of the mirror.}
  \label{freq-1}
\endminipage\hfill
\end{figure}

\begin{figure}[!htb]
\minipage{0.45\textwidth}
  \includegraphics[width=0.8\linewidth]{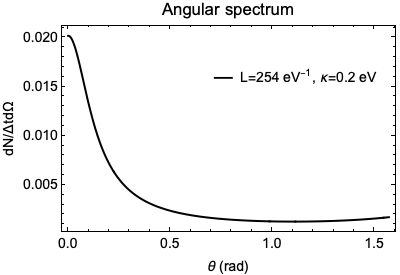}
  \caption{Angular spectrum per unit time of analog Hawking radiation.}
  \label{angular-1}
\endminipage\hfill
\end{figure}

Figures \ref{freq-1} and \ref{angular-1} illustrate the spectra for a perfect square mirror with finite area. The frequency spectrum initially grows linearly in $\omega$ in the low frequency regime as indicated by Eq.~\eqref{omega-small}. Since small frequency corresponds to long wavelength, the spectrum is simply proportional to the mirror's area. However, as the frequency gets higher, the wavelength becomes comparable to the area and diffraction emerges, leading to the wiggling in the figure. On the other hand, as indicated in Eq.~\eqref{angular-small}, for $\zeta(5)(\kappa L)^2>3\pi^2\zeta(3)$, which is the case employed in the figure, the angular spectrum has its maximum at $\theta=0$ and decreases quadratically in $\theta$ as $\theta$ gets larger.

In the recently proposed AnaBHEL collaboration \cite{AnaBHEL,Chen2017prl,Chen2020}, the flying mirror generated via laser-plasma interaction has a low reflectivity \cite{Liu2020} and the event yield per laser shot is estimated as $N\sim 0.3$ \cite{Lin2021}. Intuitively, one expects that the yield would increase as the mirror's reflectivity gets higher. However, this situation is beyond the perturbation analysis performed in \cite{Lin2020,Lin2021}. In this paper, we estimate the lower bound of the yield by only considering the normal incident modes reflected by a perfectly reflecting mirror, i.e., the actual yield should be higher if incident modes with non-vanishing transverse momenta are considered. In the AnaBHEL proposal \cite{Chen2020}, the acceleration parameter for the asymptotic Davies-Fulling trajectory is designed to be $\kappa=0.2\;\mathrm{eV}$ and the duration of the  acceleration is $\Delta t\simeq 300\mu\mathrm{m}/(3\times 10^{8}\mathrm{m/s})=1\;\mathrm{ps}= 1520\;\mathrm{eV}^{-1}$. In addition, the mirror is designed to have a transverse area of $\mathcal{A}\simeq (50\;\mu\mathrm{m})^{2}=(254\;\mathrm{eV}^{-1})^{2}$. With these parameters, the event yield per laser shot is then $N\gtrsim 0.011\times 1520\simeq 16$ events, where $0.011$ eV is the number of particles per unit time (the area under the frequency spectrum in Fig.~\ref{freq-1}). For a petawatt-class laser facility that provides 1 laser shot per minute and 8 hours of operation time per day, a 20-day experiment with an ideal detector efficiency would then give the total yield as $N_{\mathrm{total}}\gtrsim 160,000$ events, which seems promising.

\section{Conclusion}
\label{Conclusion}

The conventional approach toward radiation spectrum induced by a relativistically flying mirror in spacetimes other than (1+1) dimensions is, if possible, technically difficult. There are only limited analytic solutions: if the mirror is non-relativistic and perfectly reflecting, then it can be solved by perturbing a static mirror solution \cite{Ford1982,Neto1996,Rego2013}; if the mirror is relativistic but semitransparent, it can also be solved by perturbing a free field solution \cite{Lin2020,Lin2021}, and if the mirror is uniformly accelerated, the energy flux is studied instead \cite{candelas1977vacuum}.

To circumvent the difficulty encountered by a perfectly reflecting mirror in higher dimensional spacetimes, we begin by studying the frequency and momentum of an incident plane wave after reflection by a flying mirror instead. The result indicates that only incident waves with negligible transverse momenta can catch up the receding mirror and get reflected when the mirror is in the Davies-Fulling motion, which is relevant for mimicking Hawking radiation. In addition, since the Davies-Fulling motion covers a large portion of the acceleration phase in the AnaBHEL design, we may estimate the number of analog Hawking particles to be detected by only considering the \textit{in} modes that incident the flying mirror perpendicularly. In this situation, based on dimensional analysis, the radiation spectrum $(d^3N/d^3p)$ for an infinite area perfect mirror in (1+3)-dimensions can be determined from that $(dN/dp_{x})$ in (1+1)-dimensions, i.e., $d^3N/d^3p\propto \delta^{(2)}(\mathbf{p}_{\perp})dN/dp_{x}$. Extensions to finite mirror area and various geometries can be achieved by replacing the Dirac delta distribution by, e.g., sinc functions for a square mirror or jinc functions for a circular mirror, and the finite-size effect leads to diffraction pattern in the radiation spectrum. By adopting the parameter values designed for the AnaBHEL proposal: $\kappa=0.2$ eV, $L=254\;\mathrm{eV}^{-1}=50\;\mu\mathrm{m}$, and $\Delta t\simeq 1520\;\mathrm{eV}^{-1}= 1\;\mathrm{ps}$, the corresponding analog Hawking temperature $T_{H}\simeq 0.03$ eV falls in the far-infrared regime, and the number of analog Hawking particles per experiment (laser-shot) is $N\gtrsim 16$. In a 20-day data acquisition, we expect an ideal particle detector to detect $N_{\mathrm{total}}\gtrsim 160,000$ analog Hawking particles, which is about 50 times larger than the yield, $N_{\mathrm{total}}\sim 3,000$, for a square-root-Lorentzian (SRLD) semitransparent mirror estimated in \cite{Lin2021}.

Finally, let us comment that by striking a flying mirror with a classical plane wave and Fourier transforming the corresponding reflected wave lead to an identical radiation spectrum as that of a quantum field. However, this is only a classical signal in the Fourier space. It is only for quantum fluctuations that the Fourier components of the reflected wave modes have the interpretation as quantum particle creation from the vacuum. Therefore, it may be possible to employ classical waves in the lab to mimic the analog Hawking radiation, i.e., \textit{analog of the analog Hawking radiation}. This concept may significantly decrease the experimental difficulty of measuring the rare quantum particles. However, how to mimic the entanglement property between the analog Hawking particle and its entangled partner pair deserves further investigations.


\begin{acknowledgments}
This work is supported by ROC (Taiwan) Ministry of Science and Technology (MOST), National Center for Theoretical Sciences (NCTS), and Leung Center for Cosmology and Particle Astrophysics (LeCosPA) of National Taiwan University. K.-N. L. acknowledges supports from the Elite Doctoral Scholarship provided by NTU and NSTC, where this work was completed. 
\end{acknowledgments}

\appendix

\nocite{*}

\bibliography{main}

\end{document}